\documentclass[journal]{vgtc}

\usepackage{float}

\usepackage[bookmarks,backref=true,linkcolor=black]{hyperref} 
\hypersetup{
  pdfauthor = {},
  pdftitle = {},
  pdfsubject = {},
  pdfkeywords = {},
  colorlinks=true,
  linkcolor= black,
  citecolor= black,
  pageanchor=true,
  urlcolor = black,
  plainpages = false,
  linktocpage
}

\onlineid{326}
\vgtccategory{Research}
\vgtcinsertpkg

\manuscriptnote{}

\usepackage{amsmath,amssymb,amsthm,mathtools}

\usepackage{algpseudocode,algorithm}

\usepackage{subfig}

\usepackage{import}

\usepackage{latexsym}
\usepackage{stmaryrd}
\usepackage[noabbrev]{cleveref}
\crefformat{equation}{(#2#1#3)}

\let\V\mathbf

\def\Vdot#1{\dot{\V{#1}}}

\newcommand\Iver[1]{\llbracket#1\rrbracket}

\theoremstyle{definition}
\newtheorem{claim}{Claim}

\newtheorem{definition}{Definition}

\usepackage{newtxtext,newtxmath}
\DeclareMathAlphabet{\mathpzc}{OT1}{pzc}{m}{it}

\author{Andrew McCaleb Reach\thanks{e-mail: caleb.reach@vt.edu}\\ %
        \scriptsize Virginia Tech %
\and Chris North\thanks{e-mail: north@cs.vt.edu}\\ %
     \scriptsize Virginia Tech}

\author{Caleb Reach and Chris North}
\title{The Signals and Systems Approach to Animation}

\authorfooter{
\item
 Andrew Reach is with Virginia Tech. E-mail: caleb.reach@cs.vt.edu.
\item
 Chris North is with Virginia Tech. E-mail: north@cs.vt.edu.
}

\shortauthortitle{Reach \MakeLowercase{\textit{et al.}}: Springs,
Filters, and Transitions}

\abstract{Animation is ubiquitous in visualization systems, and a
  common technique for creating these animations is the
  \emph{transition}.  In the transition approach, animations are
  created by smoothly interpolating a visual attribute between a start
  and end value, reaching the end value after a specified duration.
  This approach works well when each transition for an attribute is
  allowed to finish before the next is triggered, but performs poorly
  when a new transition is triggered before the current transition has
  finished.  In particular, interruptions introduce velocity
  discontinuities, and frequent interruptions can slow down the
  resulting animation.  To solve these problems, we model the problem
  of animation as a signal processing problem.  In our technique,
  animations are produced by transformations of \emph{signals}, or
  functions over time.  In particular, an animation is produced by
  transforming an input signal, a function from time to target
  attribute value, into an output signal, a function from time to
  displayed attribute value.  We show that well-known
  signal-processing techniques can be applied to produce animations
  that are free from velocity discontinuities even when interrupted.}

\begin{document}
\maketitle










\section{Introduction}

Transitions \cite{stasko1990path} are widely used in information
visualization to smoothly change visual attributes, such as position,
color, opacity, size, or line width.  Transitions are useful in
situations where these attributes would otherwise be changed abruptly,
in response to, for example, new data becoming available, brushing,
dynamic queries, or navigation, and are recommended for fluid
interaction \cite{elmqvist2011fluid}.

As a simple example, imagine a visualization of a quarterly sales
dataset in which a slider allows the user to select a quarter and
sales by department for the selected quarter are shown in a bar chart.
Suppose that the goal is to animate the heights of the bars as the
selected quarter is changed.  This animation can be produced by
triggering a transition each time a new quarter is selected.  Each
transition, when triggered, smoothly interpolates between the current
height and the target height, reaching the target height after a fixed
duration has elapsed.  The specific motion of the transition is
controlled by an easing function,\footnote{The CSS animation
  specification \cite{Galineau} instead uses the term ``animation
  timing function''.} which can be used to produce animations that
begin from rest by smoothly increasing velocity and end at rest after
smoothly decreasing velocity.  However, if the user swiftly drags the
slider across many quarters, new transitions will be triggered before
the current target height has been reached.  In these cases, the new
transition will begin from rest with a zero velocity, even if the
current velocity is nonzero, causing the bar heights to abruptly
change velocity and to grow or shrink at a much slower rate than in
cases where transitions are not interrupted.  This limits the use of
animation to cases where animations are unlikely to be interrupted.
We seek to improve this animation algorithm so that it can be used
during rapid user interaction.

To produce animations that are smooth in both position and velocity,
even when interrupted, this paper introduces the signal framework for
animations.  In signal framework, animations are not produced by
triggering transitions, but instead by transforming \emph{signals},
which are functions from time to attribute values.  An animation
\emph{system} in the signal framework takes as input a signal that
maps from time to target attribute value and produces as output a
signal that maps from time to displayed attribute value.




The contributions of this paper are as follows:
\begin{enumerate}
\item It is shown in \cref{sec:transitions} that transitions behave
  poorly when interrupted.  In particular, it is shown that in the
  limit as the duration between interruptions goes to zero,
  transitions produce a constant output.
\item The signals and systems framework for animation is introduced in
  \cref{sec:signals}.  In this framework, smooth animations are
  produced by transforming signals rather than by triggering
  transitions.  This abstract technique is made concrete by techniques
  introduced in the following three sections.
\item The finite impulse response (FIR) animation technique is
  introduced in \cref{sec:fir}.  It is shown that in cases where
  transitions all have the same duration and easing function and are
  never interrupted, the the FIR technique produces the same output as
  the transition technique.  However, the FIR tech handles
  interruptions more gracefully than the transition technique for
  cases where transitions are interrupted.
\item The infinite impulse response (IIR) animation technique is
  introduced in \cref{sec:iir}.  An application of a
  spring-mass-damper system for the purposes of animation is
  presented.
\item A set of examples in which the above techniques are applied are
  presented in \cref{sec:examples}.

\end{enumerate}

\section{Transitions}\label{sec:transitions}

In this section, the transition algorithm is analyzed, both in theory
and in practice.  It is show that in cases where transitions are
interrupted, the transition algorithm fails to produce output that is
smooth in both position and velocity.


Before the transition algorithm can be analyzed, it must be defined,
and to define the transition algorithm we must first define an easing
function.
\begin{definition}[Easing function]
  A \emph{easing function} is a function $s : \mathbb R \to \mathbb R$
  from time to a real number such that $s(t) = 0$ for all $t \leq 0$
  and $s(t) = 1$ for all $t \geq d$, where $d$ is the duration of the
  easing function.
\end{definition}
This definition differs from slightly from the traditional definition
of an easing function.  Traditionally, an an easing function
traditionally ends at $t=1$.  In our definition, the easing function
is allowed to have an arbitrary duration.  The transition algorithm
can now be defined.
\begin{definition}[Transition algorithm]
  The \emph{transition algorithm} \cref{simple-transition-algorithm}
  takes an initial state $x_0 \in \mathbb R$, change requests
  $u_1,u_2,\dots,u_n$ at respective times $t_1,t_2,\dots,t_n$, and an
  easing function $S$, and produces an output function $y : \mathbb R
  \to \mathbb R$ given by
    \begin{equation}\label{eq:simple-transition}
        \begin{split}
            y_0(t) &= x_0\\
            y_i(t) &= y_{i-1}(t_i) + s(t - t_i)(x_i - y_{i-1}(t_i))\\
            y(t) &= \sum_{i=0}^n y_i(t) \Iver{t_i \leq t < t_{i+1}}
        \end{split}
    \end{equation}
    where $\Iver{p}$ is 1 if $p$ is true and $0$ if $p$ is false.
\end{definition}
We will now analyze the behavior of the transition algorithm when
interrupted and show that as the frequency of interruptions increases,
the speed of the resulting animation decreases.  This prevents the
transition algorithm from producing desirable results in cases where
interruptions are common.  Such interruptions can occur when
transition are triggered in response to interactive dynamic query
changes or in response to new data arriving.
\begin{claim}
    If the derivative of the easing function $s(t)$ vanishes at
    $t=0$ and the input is bounded, then in the limit as the
    duration between interruptions goes to zero, the output of
    the simple transition algorithm is constant.
\end{claim}
\begin{proof}
    Let $x$ be the target function, let $n$ be the number of
    samples per unit of time, and let $T=1/n$.  Our samples occur
    at $t_i = iT$ for $i$ in $1$ to $n$.  Let $y[i]$ be defined
    as $y(iT)$, and let $x[i]$ be defined similarly.  The simple
    transition algorithm is defined as
    \[
        y(t) = \sum_{i=0}^n y_i(t) \Iver{t_i \leq t < t_{i+1}}.
    \]
    Since for $t=iT$, the condition $t_k \leq t < t_{k+1}$ is
    true only for $k=i$, it follows that $y[i] = y_i[i]$.  The
    definition of $y_i[k]$ can be further simplified to
    \[
        \begin{split}
        y_i[k]
        &= y_i(kT)\\
        &= y_{i-1}(t_i) + (x_i - y_{i-1}(t_i))s(kT - t_i)\\
        &= y_{i-1}[i] + (x[i] - y_{i-1}[i])s(kT - iT)
        \end{split}
    \]
    Noting that $y_i[i] = y_{i-1}[i]$ and that $y[i+1]=y_i[i+1]$,
    we rewrite the formula as
    \[
        \begin{split}
        y[0] &= x[0]\\
        y[i+1] &= \alpha y[i] + \beta x[i]
        \end{split}
    \]
    where $\alpha = 1 - s(T)$ and $\beta = s(T)$.  This can be
    recognized as a linear discrete-time system, which has the
    solution \cite[Chapter~21]{rugh1996linear}
    \begin{equation}
        y[i] = \alpha^i x[0]
               + \sum_{k=0}^{i-1}\alpha^{i-k-1}\beta x[k]
    \end{equation}
    We now need to find $y[n]$ in the limit as $n$.  Since we
    have assumed the easing function $s(t)$ is continuous with
    zero value and derivative at $t = 0$, the first two terms of
    its Taylor series are zero, and therefore $s(T) = o(T^2)$ as
    $T \to 0$ where $o$ is the Little-O Landau symbol.  We
    therefore have
    \[
        \lim_{n\to\infty}
        \left(
        \alpha^n x[0]
        + \sum_{k=0}^{n-1}\alpha^{n-k-1}\beta x[k]
        \right)
    \]
    We will now show that the summation goes to zero and that
    $\beta^n x[0]$ goes to $x[0]$ as $n$ goes to infinity.
    Because the magnitude of $\alpha$ is near zero for
    sufficiently large $n$, $\beta$ is positive for sufficiently
    large $n$, and therefore $\beta^k$ is monotonically
    increasing with respect to $\beta$.  Since $\alpha$ is
    $o(T)$, we have that $-cT^2 < \alpha < cT^2$ for some $c$.
    Therefore $(1 - cT^2)^{n-k} < \beta < (1 + cT^2)^{n-k}$.  It
    can be easily verified that the limit of both bounds as $n$
    goes to infinity is $1$, and therefore $\beta$ must also
    approach $1$.  Since we have assumed that $u$ is bounded,
    $\sum_{k=0}^nu[k-1] = O(n)$, and therefore
    \[
        \sum_{k=0}^n \alpha \beta^{n-k} u[k-1]
        = o(T^2)O(n) = o(1/n^2)O(n) = O(1/n)
    \]
    and therefore, in the limit, goes to zero.  The term
    $\beta^nx[0]$ goes to $x[0]$ because $\beta^n$ goes to one,
    as previously discussed.
\end{proof}

\begin{algorithm}[bt]
    \caption{Simple transitions for easing function $f$}
    \label{simple-transition-algorithm}
    \begin{algorithmic}
    \State $t_1 \gets t_0$ \Comment{Time of last target change}
    \State $u_1 \gets u_0$ \Comment{Target value at $t_1$}
    \State $x_1 \gets x_0$ \Comment{Output value at $t_1$}
    \State $x \gets x_0$ \Comment{Current output value}
    \For{each frame at time $t$}
      \If{target changed to $u$}
        \State $t_1 \gets t$ \Comment{Set $t_1$ to current time}
        \State $u_1 \gets u$ \Comment{Set $u_1$ to current input value}
        \State $x_1 \gets x$ \Comment{Set $u_1$ to current output}
      \EndIf

      \State $x \gets x_1 + (u_1-x_1)f(t-t_0)$ \Comment{Set output}
    \EndFor
\end{algorithmic}

\end{algorithm}

\section{Splines}\label{sec:splines}

In this section, we analyze an alternative approach based on splines.
Splines are often used in computer graphics to smoothly animate
between user-specified key-frames.  In such systems, all key-frames
are given in advance.  By contrast, the problem considered in this
paper requires an algorithm that can process changes as they become
available.  Splines can be easily adapted to this task, and so before
presenting our technique, we first consider a solution to the
animation problem based on splines.

One approach to allowing animations to be smoothly interrupted is
to specify the easing function as a spline and modify the
velocity of the starting point on the spline to account for the
current velocity of the object.  We now analyze this approach.

A cubic Hermite spline beginning at $p_0$ with slope $\dot p_0$
at $t = 0$ and ending at $p_1$ with slope $\dot p_1$ at $t = 1$ is
given by
\[
    \begin{split}
    p(t) =
    {} &(2t^3 - 3t^2 + 1)p_0
    + (t^3 - 2t^2 + t)\dot p_0\\
    &{} + (-2t^3 + 3t^2)p_1
    + (t^3 - t^2)\dot p_1
    \end{split}
\]


\begin{definition}[Spline transition algorithm]
    The \emph{spline transition algorithm} takes an initial state
    $x_0$, initial velocity $\dot x_0$, and change
    requests $x_1,x_2,\dots,x_n$ at respective times
    $t_1,t_2,\dots,t_n$, and produces an output function
    $y : \mathbb R \to \mathbb R$ given by
    \begin{equation}\label{eq:spline-transition}
        \begin{split}
            y_0(t) &= x_0\\
            y_i(t) &= r(y_{i-1}(t_i), \dot y_{i-1}(t_i), x_i, t - t_i)\\
            y(t) &= \sum_{i=0}^n y_i(t) \Iver{t_i \leq t < t_{i+1}}
        \end{split}
    \end{equation}
    where $r$ is given by
    \[
        r(p_0, \dot p_0, p_1, t) =
        (2t^3 - 3t^2 + 1)p_0
        + (t^3 - 2t^2 + t)\dot p_0
        + (-2t^3 + 3t^2)p_1
    \]
\end{definition}
In other words, the spline transition algorithm responds to each
input change by constructing a spline segment that begins at the
current point and with the current velocity and ends $d$ seconds
in the future at the new point with zero velocity.

We now show that the spline transition algorithm successfully
approaches the target value.  Let $x$ be the target function, $n$ be
the number of samples per unit of time, and let $T = 1/n$.  Let $t_i =
iT$ and let $y[i] = y(iT)$ and $x[i] = x(iT)$.

The spline transition algorithm is defined as
\[
    y(t) = \sum_{i=0}^n y_i(t) \Iver{t_i \leq t < t_{i+1}}
\]
which simplifies as before to
\begin{align*}
  y[0] &= x_0\\
  \dot y[i+1] &= \dot r(y[i], \dot y[i], x_i, T)\\
  y[i+1] &= r(y[i], \dot y[i], x_i, T)
\end{align*}
This can be seen as a discrete-time state-space system with two
state variables: position ($y$) and velocity ($\dot y$).  Let $A$
be the matrix and $B$ be the vector such that
\[
    \begin{bmatrix}
        \dot y[k+1]\\
        y[k+1]
    \end{bmatrix} = A
    \begin{bmatrix}
        \dot y[k]\\
        y[k]
    \end{bmatrix} +
    B u[k]
\]

Let \[
    \V y[k] = \begin{bmatrix}
        \dot y[k+1]\\
        y[k+1]
    \end{bmatrix}
\]
Then $\V y[k+1] = A\V y[k] + Bu[k]$.  By making the substitution
$t = kT$ and rearranging, this can be rewritten as
\[
    \frac{\V y(t + T) - \V y(t)}{T} = \frac{A\V y(t) + Bu(t) - \V y(t)}{T}
\]
Notice that in the limit as $k$ approaches infinity, and therefore $T$
approaches zero, the left side becomes the derivative of $\V y$, and
so taking the limit of both sides gives
\[
    \Vdot y(t) = \lim_{T \to 0} \frac{A\V y(t) + Bu(t) - \V
    y(t)}{T}
    = \lim_{T \to 0} \left(
        \frac{A - I}{T} \V y(t) +
        \frac{B}{T} u(t)
    \right)
\]
The matrix $A$ is given by
\[
    A = \begin{bmatrix}
        2T^3 - 3T^2 + 1 & T^3 - 2T^2 + T\\
        6T^2 - 6T & 3T^2 - 4T + 1\\
    \end{bmatrix},
\]
and $B$ is given by
\[
    B = \begin{bmatrix}
        -2T^3 + 3T^2\\
        -6T^2 + 6T
    \end{bmatrix}
\]
Let $A'$ and $B'$ be the matrices that result from $(A - I)/T$
and $B/T$ respectively in the limit as $T$ goes to zero.  The
matrix $A'$ evaluates to
\[
    A' = \lim_{T \to 0} \frac{(A - I)}{T} =
    \lim_{T \to 0} \begin{bmatrix}
        2T^2 - 3T & T^2 - 2T + 1\\
        6T - 6 & 3T - 4\\
    \end{bmatrix} =
    \begin{bmatrix}
        0 & 1\\
        -6 & -4\\
    \end{bmatrix},
\]
and $B'$ evaluates to
\[
    B' = \lim_{T \to 0} \frac{B}{T} =
    \lim_{T \to 0}
    \begin{bmatrix}
        -2T^2 + 3T\\
        -6T + 6
    \end{bmatrix} =
    \begin{bmatrix}
        0\\
        6
    \end{bmatrix}
\]
This forms the continuous-time state-space system
\begin{align*}
  \dot x(t) &= A'x(t) + B'u(t)\\
  y(t) &= x_1(t)
\end{align*}
which has the solution
\[
    y(t) = \int_{0}^\infty e^{A(t-\tau)}B'u(\tau)\,d\tau
\]
in which the matrix exponential is used.  Evaluating this with a unit
step input $u(t)$ produces the step response.  This step response does
approach one as time increases, and so the system does eventually
reach the target value.  Interestingly, the system overshoots the
target very slightly before settling.

The spline animation technique has the advantage that interruptions do
not produce velocity discontinuities.  However, it has the
disadvantage that the designer has no creative control over the
motion of the animation.  In particular, there is no easy way to
produce motion that accelerates quickly and decelerates slowly.
The techniques presented in the following sections retain creative
control offered by the transition algorithm while allowing
interruptions without velocity discontinuities.


\section{Signals and systems}\label{sec:signals}

The techniques we propose in this paper to address the problems with
the simple transition algorithm discussed in the introduction follow
from the simple idea that the animation problem should be approached
as a problem of designing a system that transforms signals.
Fortunately, nearly all of the difficult work has been done, and we
need only to adapt well-established signal processing techniques to
our particular problem of transitions.  The field of signal processing
is vast, and there are countless ways in which to design a signal
processing system to complete a given task.  This section introduces
some of the basic concepts from signal processing, and discusses their
relevance to information visualization.

A signal is any value that changes over time.  For example, in
visualization, signals could be used to model
\begin{enumerate}
\item The mouse position
\item The mouse button state
\item Position of a data point on the screen
\item A streaming data value
\item Any computed intermediate value
\end{enumerate}

Visualizations can be seen as a function from data and program state
to an image \cite{}.  Of course, this function is in turn composed of many
smaller functions.  All of these functions may compute intermediate
values, and nearly all of these intermediate values can be regarded as
signals.  For example, a visualization might assign a position to all
nodes in a tree, and then this position might be projected to screen
space via some transformation.  Each position assigned by the tree
layout algorithm can be seen as a signal, and the screen-space
position resulting from the projection of the layout position can also
be seen as a signal.  Therefore, signal processing techniques can be
applied at any stage in the visualization pipeline
\cite{card2009information,chi1998operator}.

Systems may be classified as linear or nonlinear.  A system is
linear if it satisfies the following two properties:
\begin{enumerate}
\item If $x(t)$ is transformed to $y(t)$, then $ax(t)$ is
    transformed to $ay(t)$.
\item If $x_1(t)$ is transformed to $y_1(t)$ and $x_2(t)$ is
    transformed to $y_2(t)$, then $x_1(t) + x_2(t)$ is
    transformed to $y_1(t) + y_2(t)$.
\end{enumerate}
It is important to note that the linearity of a system is
unrelated to the linearity of its input or output signals,
i.e. both linear and nonlinear systems can transform lines to
lines, lines to curves, or curves to lines.  When using a linear
system to animate the movement of an object from one point to
another, the animation will appear to take the same amount of
time regardless of the distance between the points.  The
animation produced by a non-linear system, by contrast, may
depend on the distance between the points, or even on the
location of the points themselves.

Systems can also be classified as time-varying or time-invariant.
A time-invariant system has the property that if $x(t)$ is
transformed to $y(t)$ then $x(t - a)$ is transformed to
$y(t - a)$.  A time-varying system is any system for which this
property does not hold.  In other words, a delayed input produces
a equally delayed output for a time-invariant system, whereas a
delayed input may produce a completely different output for a
time-varying system.

This paper focuses on the class of linear, time-invariant (LTI)
systems, which is extremely well-understood.  Any LTI system can
be perfectly represented by its impulse response $h$, and given
$h$, the output $y$ for any input $x$ can be found by
\[
    y(t) = (h*x)(t)
\]
where $*$ denotes convolution, defined as
\[
    (h*x)(t) = \int_{-\infty}^\infty h(t - \tau)x(\tau)\,d\tau.
\]

Another way to classify systems is by causality.  In a
\emph{causal} system, the output at a given instant depends only
on past and present input values.  Offline systems may be
non-causal, but interactive systems such as those discussed in
this paper must be causal.

Systems may be classified as finite impulse-response (FIR) or
infinite impulse-response (IIR).  In a FIR system, the output
depends only on recent input values---all sufficiently old input
values are forgotten.  In IIR system, the output is influenced by
all past input values.  Most IIR systems encountered in practice
can be cast into a \emph{state-space} representation, in which
the output is determined by only the present input value and the
state of the system, which evolves over time in a manner
depending only on the present input at each instant.

Physical systems, such as mechanical systems composed of springs,
masses, and dampers or electrical systems composed of resistors,
inductors, and capacitors tend to be IIR.  There is a well-established
correspondence between these mechanical systems and electrical
circuits comprising capacitors, inductors, and resistors
\cite{smith2010physical} \cite{williams1996fundamentals}.  FIR
systems, by contrast, are usually associated with digital signal
processing techniques.  For animation, both FIR filters and IIR
filters can be useful.

LTI systems can be classified by the way that they affect sinusoidal
inputs.  A system that leaves low-frequency sine waves essentially
unchanged but greatly reduces the amplitude of high-frequency sine
waves is called a \emph{low-pass} filter.  Since the goal of
transitions is to smooth an input signal that changes abruptly, we
only consider low-pass filters in this paper.

We call a filter whose impulse response integrates to one
\emph{affine}.  We call an affine filter whose impulse response
is everywhere non-negative \emph{convex}.  A causal affine filter
applied to a constant input will eventually approach the input
value, and a convex system will never travel outside the range of
its input.  For the purposes of this paper, we only consider
affine filters, both convex and non-convex.

Another way to classify systems is \emph{continuous-time} or
\emph{discrete-time}.  For transitions, it's often helpful to
perform the initial design in continuous space, as it frees the
designer from having to think about the frame rate.  There are
well-known ways to convert continuous-time systems to
discrete-time systems \cite{FILTERS07}.

To design an animation, we must make design decisions.  FIR
and IIR filters provide two low-level building blocks upon which
we can construct more elaborate systems.



\section{Finite impulse response transitions}\label{sec:fir}

The first technique we propose is the FIR transition.  Like
simple transitions, FIR transitions allow the easing curve to be
specified explicitly, giving interaction designers precise
control over motion.  Unlike simple transitions, FIR transitions
maintain velocity continuity even when interrupted, and they will
make progress towards a target even if the target is continuously
changing.

\begin{definition}[Finite impulse response transition]
    The FIR transition technique gives, for input signal $x$ and
    easing function $s$, the output signal $x*\dot s$.
\end{definition}

While mathematically simple, this definition cannot be
implemented as written, since convolution involves an integral
over all time, which does not immediately imply a particular
implementation.  The convolution could, of course, be implemented
using numerical integration techniques, but this would be
needlessly wasteful for almost all cases.

In the case where the input is a step function, the math can be
simplified considerably, and is expressible using sums, which are
easy to compute, instead of integrals, which are not.  Since the
impulse response is finite, we also do not have to retain the
entire history of the program, but rather only a small window of
history.  While mathematically a FIR filter needn't be discrete-time or
have step function inputs, in practice a FIR filter is
usually implemented using discrete time.
\begin{claim}\label{fir-step}
    If $x$ is a step function that starts at $x_0$ and changes to
    $x_1,x_2,\dots,x_n$ at respective times $t_1,t_2,\dots,t_n$,
    LTI transitions simplify to
    \[
        x_0 + \sum_{i=1}^n (x_i - x_{i-1})s(t - t_i)
    \]
\end{claim}
\begin{proof}
    The input is given as
    \[
        \sum_{i=0}^n x_i \Iver{t_i \leq t < t_{i+1}}
    \]
    where $t_0 = -\infty$ and $t_{n+1} = \infty$.  The
    convolution therefore simplifies to
    \[
        \sum_{i=0}^n x_i (\dot s(t) * \Iver{t_i \leq t < t_{i+1}})
    \]
    where the convolution is taken with respect to $t$.  The
    convolution terms expand as
    \[
        \begin{split}
        \dot s(t) * \Iver{t_i \leq t < t_{i+1}}
        &= \int_{-\infty}^\infty
           \Iver{t_i \leq t - \tau < t_{i+1}} \dot s(\tau)
           \,d\tau\\
        &= \int_{-\infty}^\infty
           \Iver{t - t_i \geq \tau > t - t_{i+1}} \dot s(\tau)
           \,d\tau\\
        &= \int_{t-t_{i+1}}^{t-t_i}
           \dot s(\tau)
           \,d\tau\\
        &= s(t - t_i) - s(t - t_{i+1})
        \end{split}
    \]
    and therefore the transition can be given as
    \[
        \begin{split}
        &\sum_{i=0}^n x_i (s(t - t_i) - s(t - t_{i+1}))\\
        &= \sum_{i=0}^n x_i s(t - t_i) - \sum_{i=0}^n x_is(t - t_{i+1})\\
        &= \sum_{i=0}^n x_i s(t - t_i) - \sum_{i=1}^{n+1} x_{i-1}s(t - t_i)\\
        &= x_0 + \sum_{i=1}^n (x_i - x_{i-1}) s(t - t_i) - x_ns(t - t_{n+1})\\
        &= x_0 + \sum_{i=1}^n (x_i - x_{i-1}) s(t - t_i)
        \end{split}
    \]
    where the last step uses the fact that $t_{n+1} = \infty$ and
    therefore $s(t - t_{n+1}) = 0$.
\end{proof}
In the special case where the input is a step function and each
step in the input has a duration at least that of the easing
function, i.e. in the case where transitions are not interrupted,
the FIR transition algorithm produces results identical to those
from the simple transition algorithm.
\begin{claim}
    If each transition is allowed to finish before the next
    begins, FIR animation is equivalent to simple transitions.
\end{claim}
This claim implies that in many cases, our proposed technique is
the exact same as the simple transition algorithm.
\begin{proof}
    The simple transition algorithm is defined as
    \[
        \begin{split}
            y_0(t) &= x_0\\
            y_i(t) &= y_{i-1}(t_i) + (x_i - y_{i-1}(t_i))s(t - t_i)\\
            y(t) &= \sum_{i=0}^n y_i(t) \Iver{t_i \leq t < t_{i+1}}
        \end{split}
    \]
    The condition $t_i \leq t < t_{i+1}$ will be true for only a
    single value of $i$, and so $y(t) = y_i(t)$ for this $i$.  In
    the case where $i=0$, $y(t) = x_0$.  In the case where $i
    \neq 0$,
    \[
        y_i(t) = y_{i-1}(t_i) + (x_i - y_{i-1}(t_i))s(t - t_i)\\
    \]
    The term $y_{i-1}(t_i)$ expands as
    \[
        y_{i-1}(t_i) = y_{i-2}(t_{i-1})
                      + (x_{i-1} - y_{i-2}(t_{i-1}))s(t_i - t_{i-1})\\
    \]
    Since transitions are not interrupted, $t_i - t_{i-1}$ is
    greater than the duration of $s$, and therefore
    $s(t_i - t_{i-1}) = 1$, and thus $y_{i-1}(t_i) = x_{i-1}$.
    The definition of $y = y_i(t)$ therefore simplifies to
    \[
        y_i(t) = x_{i-1} + (x_i - x_{i-1})s(t - t_i).
    \]
    From \cref{fir-step}, FIR transitions can be defined as
    \[
        \hat y(t) = x_0 + \sum_{k=1}^n (x_k - x_{k-1})s(t - t_k).
    \]
    Clearly, in the case where $i=0$, $\hat y(t) = y(t) = x_0$.
    We now consider the case where $i \neq 0$.  Since
    $t_i \leq t$, if $k \leq i-1$, then
    $d \leq t_i - t_k \leq t - t_k$ where $d$ is the duration of
    the easing curve, and therefore $s(t - t_k) = 1$.  Similarly,
    for all $k \geq i+1$, $s(t - t_k) = 0$.  We can therefore
    expand the sum to
    \[
        \hat y(t) = x_0 +
                    \sum_{k=1}^{i-1} (x_k - x_{k-1})
                    + (x_i - x_{i-1})s(t - t_i).
    \]
    Noting that $x_{k-1}$ in the sum cancels $x_0$ for $k=1$ and
    that $x_k$ is canceled by $x_{(k+1)-1}$ for all $k < i-1$,
    the sum further simplifies to
    \[
        \hat y(t) = x_{i-1} + (x_i - x_{i-1})s(t - t_i),
    \]
    which is clearly equal to $y(t)$.
\end{proof}

\begin{algorithm}[bt]
    \caption{Continuous-time FIR transitions for step input and
    easing function $f$.}
    \label{fir-step-alg}
    \begin{algorithmic}
    \State $\textrm{value} \gets \textrm{initial value}$
    \State $\textrm{targets} \gets \textrm{Queue}()$\\
    \For{each frame at time $t$}
      \If{target changed to $u$}
        \State $\textrm{targets.enqueue}(\textrm{Target}(\textrm{time: } t, \textrm{value: } u))$
      \EndIf\\

      \While{targets not empty and $\textrm{targets.peek}()\textrm{.time} < t - d$}
        \State $\textrm{value} \gets \textrm{targets.dequeue}()\textrm{.value}$
      \EndWhile\\

      \State $\textrm{out} \gets \text{value}$
      \State $\textrm{prev} \gets \text{value}$
      \For{each target in targets (in chronological order)}
        \State $\textrm{out} \gets \textrm{out} + f(t - \textrm{target.time})(\textrm{target.value} - \textrm{last})$
        \State $\textrm{prev} \gets \text{target.value}$
      \EndFor
    \EndFor
\end{algorithmic}

\end{algorithm}

\begin{algorithm}[bt]
    \caption{Discrete-time FIR algorithm.}
    \label{fir-discrete}
    \begin{algorithmic}
    \State $\textrm{buffer} \gets \textrm{Array}(\text{size: } n \text{, initial: } 0)$
    \State $\textrm{index} \gets 0$
    \For{each input value $u$}
      \State $\textrm{buffer}[\textrm{index}] \gets u$\\

      \State $\textrm{out} \gets 0$
      \State $k \gets \textrm{index}$
      \For{$i$ in 0 to $n - 1$}
        \State $\textrm{out} \gets \textrm{out} +
               \textrm{coefs}[i] \cdot \textrm{buffer}[k]$
        \State $k \gets k - 1$
        \If{$k < 0$}
          \State $k \gets n - 1$
        \EndIf
      \EndFor\\

      \State $\textrm{index} \gets \textrm{index} + 1$
      \If{$\textrm{index} > n - 1$}
        \State $\textrm{index} = 0$
      \EndIf
    \EndFor
\end{algorithmic}

\end{algorithm}

An implementation of continuous-time FIR animation is shown in
\cref{fir-step-alg}.  An implementation of the discrete-time FIR animation
algorithm is shown in \cref{fir-discrete}.  These algorithms can be
trivially extended to operate on vector input signals to produce 2D or
3D animations by means of a matrix easing function: if the easing
function maps from time to a matrix and the input is vector-valued,
then the algorithm works as written.  The key difference between these
algorithms and the simple transition algorithm is that these
algorithms retain a queue of in-progress transitions whereas the
simple transition algorithm only tracks a single transition.



\section{Infinite impulse response transitions}\label{sec:iir}

\begin{figure}
  \centering
  \includegraphics{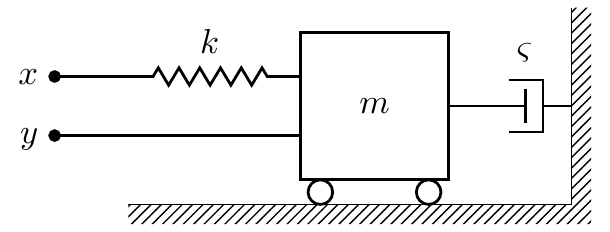}
  \caption{A mass spring damper system, with spring constant $k$,
  damper constant $\varsigma$, and mass $m$.  The input to the
  system is given as a horizontal position $x$, and the output is
  provided as a horizontal position $y$.}
 \label{spring-mass-damper}
\end{figure}

The second technique we propose is the IIR animation.  Unlike FIR
transitions, IIR animations do not take an explicit easing curve
as a parameter.  Instead, IIR systems are based on differential
equations, and can be used to model mechanical and electrical
systems that are linear and time-invariant.
\begin{definition}
    An IIR animation is given by
    \begin{equation}\label{iir-state-space}
        \begin{split}
            \dot x(t) &= Ax(t) + Bu(t)\\
            y(t) &= Cx(t) + Du(t)
        \end{split}
    \end{equation}
    where $u(t)$ is the input vector, $x(t)$ is the state vector,
    $A$, $B$, $C$, and $D$ are matrices and $y(t)$ is the output
    vector.
\end{definition}
The formula above is simply the well-known state-space
representation of a linear system.  Such systems can be used to
model physical mechanical systems of springs, masses, and
dampers, as well as circuits comprising resistors, capacitors,
and inductors.  Such systems have natural-looking behavior, and
are therefore useful for producing natural-looking animations in
visualization.  An example system is shown in
\cref{spring-mass-damper}.  For this system, the equations of
motion are given by
\[
    m\ddot x = k(u - x) - \varsigma \dot x.
\]
Putting this equation into state-space form requires introducing
an additional state variable to capture the second derivative of
$x$.  By setting, $q = \dot x$, the above equation can be written
as
\[
    \begin{split}
    \dot q &= \frac{k}{m}(u - x) - \frac{\varsigma}{m} \dot x\\
    \dot x &= q
    \end{split}
\]
which is easily put into the form given by \cref{iir-state-space}
using state vector $(x,q)$.  More detailed examples can be found
in most signal processing textbooks,
e.g. \cite{smith2010physical} \cite{oppenheim1983signals}
\cite{rugh1996linear}.  These systems can produce fast-in,
slow-out animations, as well as spring-type effects where the
object wiggles back and forth as it settles.
Fast-in\slash{}slow-out animations, i.e. animations that
accelerate quickly from rest and decelerate slowly, are
recommended by Google's Material Design guidelines
\cite{materialdesign}.  Note that in this paper,
fast-in\slash{}slow-out refers to an animation that smoothly
accelerates from rest and decelerates smoothly to rest, but that
attains its maximum speed before the animation is halfway
finished.  Some papers \cite{dragicevic2011temporal} instead use
fast-in to refer to an animation that begins with a nonzero
velocity.

\begin{figure}
    \centering
    \includegraphics{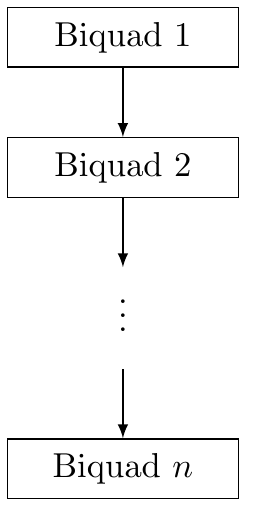}
    \caption{An order $n$ IIR filter can be implemented as a
    series combination of $\lceil n/2 \rceil$ biquad (order 2
    IIR) filters.}
    \label{fig:biquads}
\end{figure}

The trajectory of such systems can be animated using well-known
numerical integration techniques, e.g. Runge-Kutta or Verlet
integration.  Another option, which we advocate, is converting
the continuous-time system to a discrete-time system, which can
be accomplished using well-known methods such as the
impulse-invariant method, the bilinear transform, or the matched
z transform \cite{smith2010physical}.  When converting systems in
this manner, high-order systems (i.e. systems with
high-dimensiona state vectors) are usually broken up into
\emph{biquads}, which are order-two IIR filters (i.e. IIR filters
with two state variables).  One possible implementation of a
biquad is shown in \cref{iir-discrete}, and an implementation of
a high-order system in terms of biquads placed in series is shown
in \cref{fig:biquads}.


\begin{algorithm}[bt]
\caption{Discrete-time IIR realized via the Transposed Direct Form II \cite{FILTERS07} with
coefficients $a_i$ and $b_i$.}
\label{iir-discrete}
\begin{algorithmic}
    \For{each input value $x$}
      \State{$y \gets s_1 + b_0x$} \Comment{Compute output}
      \State{$s_1 \gets s_2 + b_1x - a_1y$} \Comment{Update state}
      \State{$s_2 \gets b_2x - a_2y$}
    \EndFor
\end{algorithmic}

\end{algorithm}

\section{System diagrams}

The preceding two sections discussed two low-level techniques for
creating animations.  This section discusses system diagrams, which
are used in signal processing to represent higher-level systems that
are assembled out of simpler components.  In a system diagram, blocks
represent components (subsystems) and wires represent signals.  Each
block has an associated state that changes over time, and the state of
the composite system is given by the combination of the states of the
components.  Each block has state and behavior, and is therefore similar to
the notion of an object in object-oriented programming.

If all components in the system are LTI, then the composite system is
also LTI.  If two LTI components are connected in series, then
resulting impulse response is the convolution of the impulse responses
of the components.  If the first component transforms an input sine
wave by multiplying the amplitude by $a_1$ and delaying the output by
$\tau_1$, and the second component transforms the same input sine wave
by multiplying the amplitude by $a_2$ and delaying the output by
$\tau_2$, then a connecting these two components in series results in
a system that multiplies the amplitude of the input sine wave by
$a_1a_2$ and delays the output by $\tau_1 + \tau_2$.  If two LTI
components are connected in parallel, then the impulse responses add.
Connecting several LTI components in series, each of which smooths its
input a moderate amount, can be used to produce a system that smooths
its input heavily.




\section{Applications}\label{sec:examples}

We now discuss several example applications of these ideas as a
way to illustrate the signals and systems design process.  These
examples represent scenarios that involve frequent interruptions
during animations.



\subsection{Animating histogram bins}

Imagine a dynamic query \cite{ahlberg1992dynamic} or scented widget
\cite{willett2007scented} system in which there is one histogram per
data attribute and each histogram has a selection control that can be
used to filter along that dimension.  For example, crossfilter
\cite{crossfilter} provides this type of dynamic query.  As the user
drags the selection control, points will enter and exit the selection,
causing the heights of bins in the other histograms to change
abruptly.  Our goal is to smoothly animate the heights of bins
changing.

\begin{figure}
    \centering
    \subfloat[b][Filter height]{\includegraphics{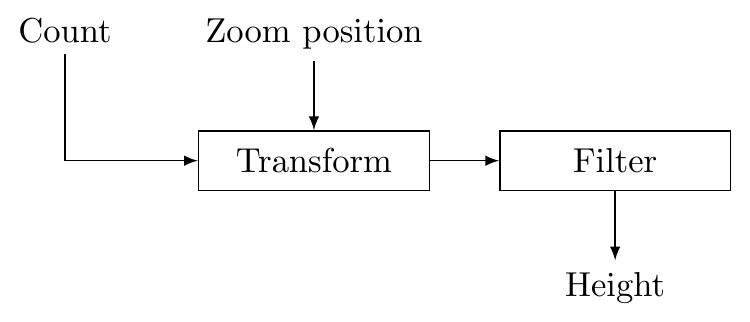}}\vspace{1em}
    \subfloat[a][Filter count]{\includegraphics{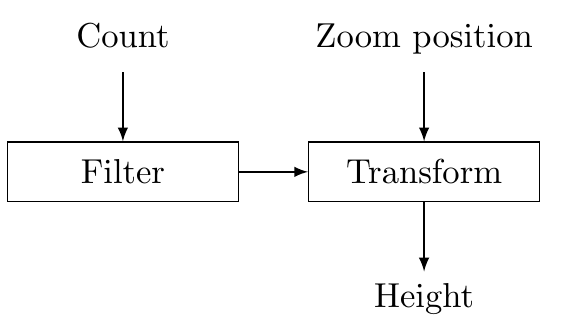}}
    \caption{If the heights of histogram bins are smoothed (a),
      zooming is not immediate, which can be problematic if a pinch
      gesture is used to zoom.  Therefore, the counts should be
      smoothed before projecting counts to heights (b).}
  \label{fig:hist-bins}
\end{figure}

The height of each bin can be regarded as a signal, and this signal
could be smoothed using the methods discussed in this paper.  However,
suppose that we wish to also allow the user to zoom in on the $y$ axis
using a pinch gesture.  Since the height of a bar on the screen
depends on both the count and the zoom position, if a smoothing filter
were applied to the height of the bar, it would also smooth the zoom
gesture.  Since the zoom gesture is already continuous, this smoothing
would introduce an undesirable disconnect between the user's actions
and the on-screen response.  In order to avoid this, the smoothing
filter should be applied to the bin count signal, and the height
should then be computed based on the bin count signal and the zoom
position signal.  Smoothing the count of each bin produces the same
on-screen result as smoothing its height in the case where the $y$
axis is static, but it allows for immediate changes to zoom position.
These two approaches are shown in \cref{fig:hist-bins}.

\subsection{Animating permutations}\label{sec:permute}

Imagine animating a permutation of a collection of objects arranged in
a row.  If each object is animated along a line to its new
destination, then objects will often collide during the course of the
animation.  If each object is instead animated along an elliptical
trajectory, this tendency will be significantly reduced, and the paths
of objects will form an arc diagram \cite{heer2010tour}.  In this
section, we describe a LTI FIR filter that has this behavior.

Since each input point lies on a line, the input to the system is
a one-dimensional vector, i.e. a scalar.  The output, however, is
a two-dimensional vector.  Therefore, the step response for this
system is a time-varying two-by-one matrix, i.e. a time-varying
vector.  To qualify as an easing function, the step response must
begin at $(0,0)$ and end at $(1,0)$.  An ellipse with a
horizontal major axis having these two endpoints is given by
\begin{equation}\label{ellipse-theta}
    \V x(\theta) = \frac{1}{2}\begin{bmatrix}
            \cos(\theta) + 1\\
            \alpha\sin(\theta)
    \end{bmatrix}
\end{equation}
where $\alpha$ is the aspect ratio of the ellipse.  At
$\theta = \pi$, this simplifies $(0,0)$, and at $\theta = 2\pi$,
this simplifies to $(1,0)$.  To use this trajectory as an easing
function, it must be parameterized by time instead of by
$\theta$.  A tempting choice is to set $\theta = \pi + \pi t$,
i.e. linearly interpolate $\theta$ between the beginning and
ending angles.  When $\alpha = 1$, the object travels a circular
trajectory at constant speed.  When $\alpha \neq 1$, however, the
object travels an elliptical trajectory at a time-varying speed.

To avoid this coupling between the shape of the ellipse and the
speed of the object, a constant-speed parameterization must be
found.  The arc length of a segment of \cref{ellipse-theta} from
$\theta_1$ to $\theta_2$ is given by
\[
    s(\theta_1,\theta_2) = \int_{\theta_1}^{\theta_2} \|\Vdot x(\theta)\|\,d\theta
\]
where $\Vdot x$ denotes the derivative of $\V x$.  Since the
trajectory begins at $\pi$ and ends at $2\pi$, the arc-length of
the entire trajectory is given by $s(\pi,2\pi)$.  The distance
traveled from $\pi$ to some $\theta$ between $\pi$ and $2\pi$
is given by $s(\pi,\theta)$.  The portion of the total distance
traveled at a given $\theta$ is therefore given by
\[
    \sigma(\theta) = \frac{s(\pi, \theta)}{s(\pi,2\pi)}.
\]
If $t = \sigma(\theta)$, then the trajectory has constant speed.
The complete trajectory can then be given using the inverse of
$\sigma$ as $\V x(\sigma^{-1}(t))$.

We now turn our attention to the computation of $\sigma^{-1}$.
Since $\sigma$ is defined in terms of $s$, which is in turn
defined in terms $\Vdot x$, we first find $\Vdot x$:
\[
    \Vdot x =
    \frac{1}{2}\begin{bmatrix}
        -\sin \theta\\
        a \cos \theta
    \end{bmatrix}.
\]
Therefore,
\[
    \|\Vdot x\| = \sqrt{\sin^2 \theta + a^2 \cos^2 \theta}
\]
and so
\[
    s(\theta_1,\theta_2) =
    \frac{1}{2}
    \int_{\theta_1}^{\theta_2} \sqrt{\sin^2 \theta + a^2 \cos^2 \theta}\,d\theta
\]
This integral does not, in general, have a closed-form solution.
However, it can be rewritten in terms of the incomplete elliptic
integral of the second kind, $E(\varphi, k)$, defined as
\[
    E(\varphi \mid k^2) = \int_0^\varphi \sqrt{1 - k^2 \sin^2 \theta}\,d\theta,
\]
by rearranging $s$ as follows:
\[
    \begin{split}
    s(\theta_1,\theta_2)
    &=
    \frac{1}{2}
    \int_{\theta_1}^{\theta_2}
    \sqrt{\sin^2 \theta + a^2 \cos^2 \theta}
    \,d\theta\\
    &=
    \frac{1}{2}a
    \int_{\theta_1}^{\theta_2}
    \sqrt{\frac{1}{a^2}\sin^2 \theta + \cos^2 \theta}
    \,d\theta\\
    &=
    \frac{1}{2}a
    \int_{\theta_1}^{\theta_2}
    \sqrt{\frac{1}{a^2}\sin^2 \theta + 1 - \sin^2 \theta}
    \,d\theta\\
    &=
    \frac{1}{2}a
    \int_{\theta_1}^{\theta_2}
    \sqrt{1 - \left(1 - \frac{1}{a^2}\right)\sin^2 \theta}
    \,d\theta\\
    &= \frac{1}{2}a \bigg(E(\theta_2 \mid k^2) - E(\theta_1 \mid k^2)\bigg)
    \end{split}
\]
where $k^2 = 1 - (1/a^2)$.

We have now specified $s$, and therefore $\sigma$, in terms of
$E$.  Implementations of $E$ can be found in many commonly-used
mathematical libraries.  In order to compute the trajectory,
however, we must resort to numerical methods to find the inverse
$\sigma^{-1}$.  We now have everything we need to compute
$x(\sigma^{-1}(t))$, which gives movement along the ellipse at a
constant rate.

Velocity discontinuities can be avoided by using an extra easing
function.  Instead of computing the easing function $f$ in
\cref{fir-step-alg} by $f = (x \circ \sigma^{-1})(t)$, we can compute
$f = (x \circ \sigma^{-1} \circ g)(t)$, where $g$ is a scalar easing
function.  Given that our method for computing this function relies on
running a root finding algorithm on a function involving elliptic
integrals, this function is almost certainly costly to evaluate.  In
the discrete-time case, the solution is simple: compute the FIR filter
coefficients by evaluating this function at a fixed set of points.
This is the so-called \emph{impulse invariant} method for discretizing
a continuous-time system.  These coefficients need only be generated
once, and once generated, the output at each frame can be computed
using one multiplication and addition per coefficient.  In the
continuous-time case, we can again compute the function at fixed
points, and then interpolate between the samples to reconstruct the
continuous function.

To complete this example, we discuss a practical example of the inner
easing function $g$.  One choice is to model $g$ on the step response
of an IIR system.  Consider the system for a one-pole filter
\[
    \dot x(t) = x(t) + a(u(t) - x(t)).
\]
The impulse response of this system decays exponentially.  Used alone,
this filter would not smooth the input sufficiently for animation use.
Therefore, we form a four-pole filter by cascading four one-pole
filters, each given by the above equation.  The step response of this
four-pole filter can be then be used as the easing function $g$, and
then $f = (x \circ \sigma^{-1} \circ g)(t)$ can be used in
\cref{fir-step-alg}.

\begin{figure}
    \centering
    \includegraphics{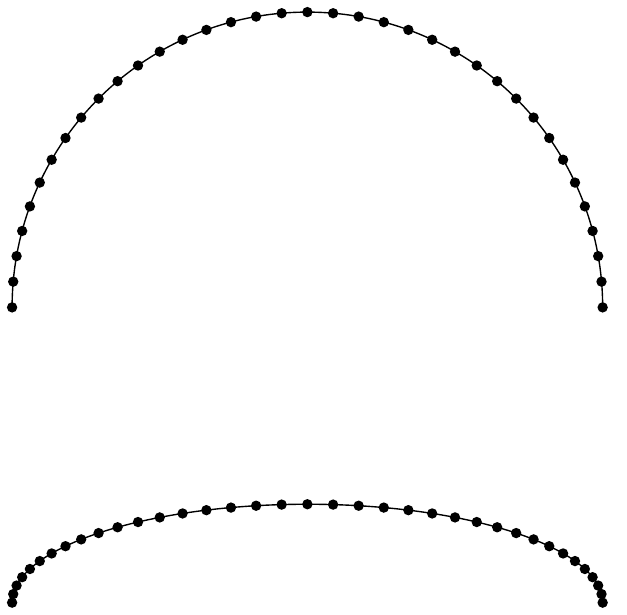}
    \caption{Constant-speed movement along a circular arc can be
    easily achieved by changing the angle at a constant rate.
    However, if the resulting points are scaled so that arc
    becomes elliptical rather than circular, the resulting motion
    is slower near the endpoints of the major axis and faster
    near the endpoints of the minor axis.}
\end{figure}



\subsection{Animating changes to a text document}

In this section we discuss the problem of animating the history of a
text document in a manner similar to the Diffamation system
\cite{chevalier2010using}.  An important contrast between our
technique and Diffamation is the behavior of the revision selector: in
Diffamation, the revision selector is treated as a playback slider for
the entire animation, whereas in ours, the animation is dynamically
generated using the selected revision as a target.  As a consequence,
dragging the selector quickly in Diffamation produces a sped-up
animation, and so a separate interaction technique is used to compare
distance revisions, whereas with our technique, moving the slider
quickly does not cause objects on screen to move excessively fast.  A
similar tool is Gliimpse \cite{dragicevic2011gliimpse}, which animates
back and forth between markup code and the rendered document.

Let the text document be modeled mathematically as a sequence
$c_1,c_2,\dots,c_n$ of characters that exist in the document's
history.  Let $p_i(t)$ be true if $c_i$ is present in the
document at time $t$, and let $\V x_i(t)$ give the on-screen
position of $c_i$ at time $t$.  The notation $\V x_i$ denotes the
vector at index $i$ in a sequence of vectors, not the
$i^{\text{th}}$ element of vector $\V x$.  The document can then
be drawn at time $t$ by drawing each $c_i$ for which $p_i(t)$ is
true at position $\V x_i(t)$.  Consider an interface in which the
user can control the revision displayed using a slider.  We seek
an animation technique that allows the user to easily see where
changes have occurred.

There are three types of abrupt changes that occur in the
animation-free visualization:
\begin{enumerate}
\item Characters appear abruptly at times when $p_i(t)$ changes
    from false to true.
\item Characters move abruptly at times when $\V x_i(t)$ changes
    abruptly.
\item Characters disappear abruptly at times when $p_i(t)$
    changes from true to false.
\end{enumerate}

Suppose we transition characters in by gradually enlarging them
from an infinitesimal size to their normal size.  At they are
enlarged, their color slowly changes from blue to black to show
that they are being added.  When their position changes, they
should move smoothly to their new position.  And when they are
deleted, they should slowly shrink to the infinitesimal size
before disappearing completely, with color changing to red to
show that they are being deleted.

We first construct the signal $\Iver{p_i(t)}$, which is one when
$p_i(t)$ is true and zero otherwise.  Next, we define the signal
$\psi_i$ as a smoothed version of $\Iver{p_i(t)}$.  Now $\psi_i$
can be used directly to control the size of the character.
Similarly, let $\V y_i$ be defined as a smooth version of
$\V x_i$, and can be used to directly control the position of the
character.  Let $\chi$ be defined as $\Iver{p_i(t)}$ filtered by
a system having a step response that is smooth and compactly
supported.

We have not yet defined what the position $\V x(t)$ of a
character should be in the case where it hasn't yet appeared in
the document or if it has been removed from the document.  To
specify this, we first define a total ordering on all characters
in the document.  We start by defining a partial ordering $<$
between characters in a document.  We assume that characters can
be added and deleted but not moved or reinserted.  We say that
$c_i < c_j$ if there exists some time $t$ such that $p_i(t)$ and
$p_j(t)$ are both true and $c_i$ appears earlier in the document
than $c_j$ at time $t$.  We now extend this partial order to a
total order.  If the ordering of $c_i$ and $c_j$ is left
unspecified by the partial order, then the set of times for which
$p_i(t)$ is true must not overlap the set of times for which
$p_j(t)$ is true, otherwise the partial order would be defined.
Therefore, we let $c_i < c_j$ if $c_i$ is inserted at an earlier
time than $c_j$, and $c_j < c_i$ if $c_j$ is inserted at an
earlier time than $c_i$.

\begin{figure*}
    \centering
    \includegraphics{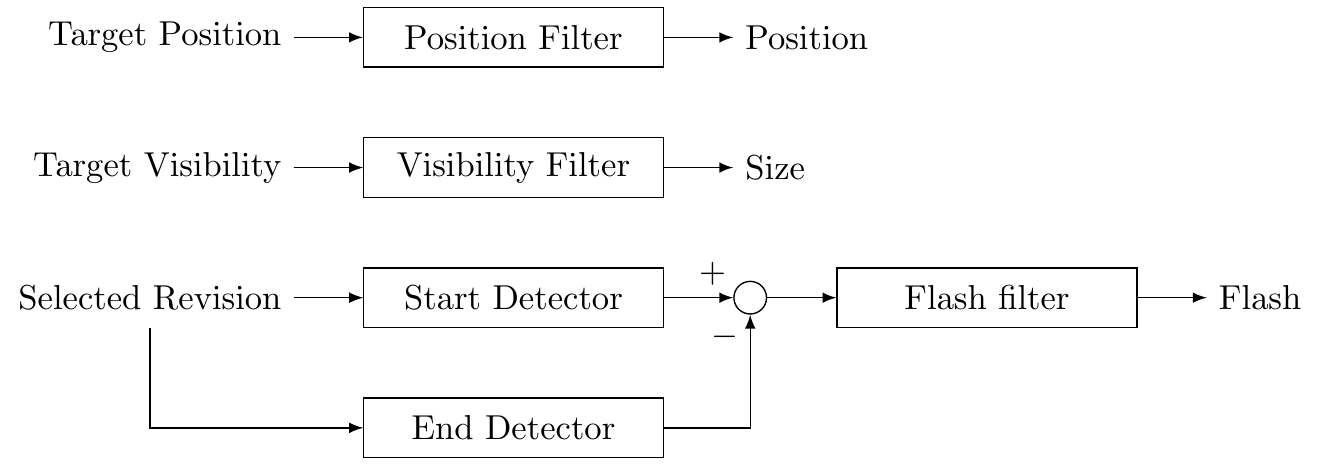}
    \caption{System for animating each text character.  The filters
      used can either be IIR as given by
      \cref{iir-discrete} or FIR filter as given by
      \cref{fir-step-alg,fir-discrete}.}
    \label{fig:text-sys}
\end{figure*}

We can now define the position $\V x_i(t)$ of a character that is
not currently visible.  Let $c_j$ be the least (in the sense of
the total ordering) currently visible character such that
$c_i < c_j$.  Then $\V x_i(t)$ is located at position of $c_j$.
Assume that the document contains an invisible end-of-file
character that is always present, and therefore such a $c_j$
always exists.  The total order of characters in an example
document is shown in \cref{text-doc}.  The signal processing
diagram for processing a character is shown in \cref{fig:text-sys}.

\begin{figure*}
\input{text}
\caption{Animating changes to a single-line document.  Each line
above represents a revision to the document.  Each column
represents a character that exists in the document from some
start time to some end time.  A natural partial order exists
between characters that exist in the document at the same time.
To define the total ordering, shown above as the ordering of the
columns, we must further specify that new text inserted in the
same location as deleted text should be ordered before the
deleted text.}
\label{text-doc}
\end{figure*}

\section{Discussion}

We have now introduced two low-level building blocks for
animations, FIR animations and IIR animations, and a way to
combine these low-level building blocks together to produce high-level
functionality, the system diagram.  Using this approach,
animations can be produced that are smooth, even in the presence
of interruptions.

Algorithms for implementing discrete-time FIR and IIR filters are
well-known, and can be found in most textbooks on digital signal
processing, e.g. \cite{FILTERS07}.  In particular, both
\cref{iir-discrete} and \cref{fir-discrete} are well-known
algorithms.  The implementation given by \cref{fir-step-alg} is a
simple implementation of a FIR filter applied to a step-function
input, and is presumably the same algorithm that Apple has used
to implement additive animation \cite{additiveanimation}.  One
disadvantage of FIR animations when compared with IIR animations
are that FIR filters require space and computational time
proportional to the duration of the animation.  IIR animations,
on the other hand, can be made arbitrarily long with constant
space and time complexity.

For an existing code base, the easiest way to switch from simple
transitions to LTI transitions is to change the algorithm to use
\cref{fir-step-alg}.  In practice, this means maintaining a list of
in-flight transitions rather than a single transition.

Since the frame-rate is usually constant in visualization
systems, we recommend using a discrete-time system whenever
possible.  However, it is often easier to think in terms of
continuous time, as this allows a system to be designed without
paying attention to the specific frame rate.  In other words, the
system should have some essential behavior that should exist
regardless of the rate at which it is sliced into frames.
Therefore, we recommend designing the system in continuous time
and then transforming the continuous-time system into a
discretized version.  For a bandlimited FIR system, this can be
done by the impulse invariant method, which simply samples the
impulse response.  For an IIR system, this can be done using
several means.  The impulse-invariant method converts the
differential equations to difference equations that trace the
same path but at discrete points in time.  The disadvantage of
the impulse-invariant method is aliasing, although this will not
be especially problematic for filters used for animations, as
they will have low-pass characteristics.  The bilinear transform
is another technique for transforming a discrete-time system into
a continuous-time system.  The bilinear transform does not
introduce aliasing, but it does warp the frequency response of
the continuous-domain filter.  The matched z-transform is still
another technique for converting from a continuous-time system to
a discrete-time system.  All of these techniques are covered in
many digital signal processing textbooks,
e.g. \cite{smith2010physical}.

Another advantage of our technique is what might be called an
\emph{emergent interaction}---an interaction that arises
serendipitously from the combination of independent features.  By
combining dynamic queries with LTI animation, the interface
affords a new interaction technique: the user can, by wiggling
the query region rapidly over a range of positions, see the
average result for queries within this range of positions.

This paper has focused on linear, time-invariant animations.
There is also a rich theory for linear, time-varying systems,
although such systems tend to be more difficult to analyze, as
commonly-used tools for analyzing LTI systems, such as the
Fourier, Laplace, and z transforms, no longer apply.

For FIR transitions, the choice of easing function makes a large
impact on the aesthetics of the motion.  If symmetric motion is
desired, B-spline window functions \cite{toraichi1989window} provide a
natural solution, and are equivalent to box filters connected in
series.  If fast-in slow-out motion is desired, as we have
recommended, a series combination of one-pole filters, as described in
\cref{sec:permute}, is an attractive alternative.



\section{Related work}

Many systems exist that use animation to help users comprehend changes
\cite{%
  bostock2011d3,%
  heer2007animated,%
  stasko1990path,%
  baecker1969picture,%
  purchase2006important,%
  yee2001animated,%
  bach2014graphdiaries,%
  plaisant2002spacetree,%
  chang1995animation,%
  thomas2001applying,%
  rosling2009gapminder%
}.  The problem we consider in particular, that of interrupted
transitions, has been previously discussed.  The W3 CSS animation
specification \cite{Galineau} mentions interrupted animations and
recommends reversing the animation for the specific case where the new
value for the new transition equals the old value of the current
transition.  This technique results in an animation with a velocity
discontinuity and symmetry about the time $t_1$ at which the animation
is reversed.  For this case, our technique produces an animation with
no velocity discontinuities and symmetry about $t_1 + d/2$.  Note that
if we want an asymmetric transition for a button such that the
transition from the normal state to the hover state and the transition
from the hover state to the normal state are reverses of one another,
we can simply use a symmetric easing function and then a nonlinear
function to the tween signal.  For example, squaring the tween signal
will produce such an asymmetry.

Apple's additive animation solves the interruption problem by
adding the transitions together.  In the case where the same
easing curve is used for each transition, this is equivalent to
FIR transitions.  The case where transitions are allowed to
change is equivalent to a linear, time-varying finite impulse
response system.  However, changing the easing curve can lead to
problematic animations.  In the LTI case, if the step response of
a one-input one-output system is bounded between zero and one,
then the output will always be a convex combination of points
from the input signal---i.e. it will never overshoot the target.
If the easing curve is allowed to vary, however, this is no
longer true, even for bounded easing curves.  Consider, for
example, an animation from point $A$ to point $B$ using an easing
curve with a long duration, quickly interrupted by an animation
back to $A$ using an easing curve with a short duration.  The
resulting animation will drastically overshoot $A$ and then
slowly move back.

Many uses of animation in visualization and computer graphics fall
outside the scope of this paper.  For example, there are many
techniques in which complex animations arise from interactions between
simple particles, e.g. dust and magnets \cite{yi2005dust}, force
directed graph layout \cite{fruchterman1991graph}, and boids
\cite{reynolds1987flocks}.  In game design, \emph{motion blending}
\cite{feng2012analysis,kovar2003flexible,mukai2005geostatistical}
techniques are often used to create new animation clips by blending
together two or more preexisting animation clips.  All of these
techniques do not attempt to solve the problem that we consider, which
is animating an attribute to a target value that may change over time.


The field of signal processing has produced rich body of knowledge
about filter design, and there are many excellent references
available, e.g. \cite{oppenheim1983signals,FILTERS07,rugh1996linear}.
Note that our FIR implementation differs from the usual implementation
of digital filters.  Digital filters usually operate on a discrete,
bandlimited, uniformly sampled signal.  Our filters instead operate
directly on the continuous attribute signal.  We don't expect the
attribute signal to oscillate, so aliasing isn't as problematic as it
is in classical digital signal processing.  We also don't use an ideal
brickwall lowpass filter, as we wish to avoid the ringing associated
with the sinc function.

The field of \emph{control theory} \cite{nise2014control} has also
produced a rich body of results that may be applied to animation.  For
example, if on-screen objects are modeled as propulsion vehicles, then
optimal control theory could be used to solve the problem of animating
these objects from one point to another with the least amount of fuel
used.  While such animations may be useful for visualization, the
difficulty of solving optimal control problems makes this approach
unappealing in practice.  A related field in robotics is \emph{motion
  planning} \cite{latombe2012robot}, which concerns the problem of
controlling navigating a robot from one location to another.  Like
control theory, motion planning approaches are likely needlessly
complex for the problem of animation.


Our technique is somewhat related to kernel smoothing
\cite{Hastie2011}.  A key difference is that kernel smoothing deals
with discrete points, whereas our method deals with a continuous step
signal.  Also, in kernel smoothing, the output at a given point in
time depends on inputs occurring in the future, thus giving a
non-causal filter and precluding a realtime implementation.  Moreover,
some kernels commonly used in kernel smoothing (eg. the Gaussian
kernel) are everywhere nonzero and thus cannot be used for LTI
transitions (although they could, of course, be truncated in
practice).

Functional reactive programming
\cite{cooper2006embedding,courtney2003yampa,elliott1997functional,czaplicki2013asynchronous,bainomugisha2012survey}
offers a programming interface in which signals are manipulated as
first-class values.  Functional reactive programming systems are an
area of active research, and could provide a promising method of
allowing LTI animation to be effortlessly integrated at any step in a
visualization pipeline.










\section{Conclusion}

This paper has presented LTI animation, an animation model, based
on well-known techniques from signal processing, that responds
well to rapid changes in attribute values.  We have shown that
for cases where attribute values do not change during
transitions, our method is equivalent to the transition model,
but that our approach handles rapidly-changing signals gracefully
and without the bizarre effects of the transition model.

The primary insight behind our method is that animations should not be
constructed to smoothly connect one state to another: rather,
animations should take into account a the history of states and the
current target state.  From this perspective, animations can be seen
as a transformation from an input signal that describes the current
target over time to a smoothed output signal.  In particular, an
easing function can be seen as a step response of a linear,
time-invariant filter that describes this transformation.

We have also introduced several algorithms for computing the LTI
animation convolution.  The FIR step method is straightforward and is
somewhat similar to FIR filter implementations.  It is likely the best
choice for most real-world usages.  We note that these techniques are
largely interchangeable: in practice, the impulse response for an IIR
filter can be truncated and then used as the impulse response for a
FIR system.  These systems will then produce the same results modulo
truncation error, which can be made arbitrarily small.

Future work might consider an extension of this technique to staggered
transitions; however, there is evidence \cite{chevalier2014not} that
staggered transitions provide little perceptual benefit.  Future work
might also consider extensions to zooming and panning animations
\cite{van2003smooth,van2004model} for which considerable care is
required to produce efficient trajectories.

\acknowledgments{This research was partially supported by NSF grant
  IIS-1527453.}



\bibliographystyle{abbrv}
\bibliography{refs.bib}
\end{document}